\documentclass{desyproc}

\begin{document}
\title{Comissioning of TREX-DM, a low background Micromegas-based Time Projection Chamber for low mass WIMP detection}
\author{{\slshape Francisco~Jos\'e~Iguaz\footnote{Corresponding author (iguaz@unizar.es)}, Javier~Garc\'ia~Garza,
Francisco~Aznar\footnote{Present address: Centro Universitario de la Defensa, Universidad de Zaragoza, Spain.},
Juan~Francisco~Castel, Susana~Cebri\'an, Theopisti~Dafni, Juan~Antonio~Garc\'ia,
Igor~Garc\'ia~Irastorza, \'Angel~Lagraba, Gloria~Luz\'on, Alberto~Peir\'o}\\[1ex]
$^1$Laboratorio de F\'isica Nuclear y Astropart\'iculas, Universidad de Zaragoza, Spain}

\contribID{iguaz\_francisco}

\confID{11832}  
\desyproc{DESY-PROC-2015-02}
\acronym{Patras 2015} 
\doi  

\maketitle

\begin{abstract}
Dark Matter experiments are recently focusing their detection techniques in low-mass WIMPs,
which requires the use of light elements and low energy threshold. In this context,
we describe the TREX-DM experiment, a low background Micromegas-based Time Projection Chamber for low-mass WIMP detection.
Its main goal is the operation of an active detection mass $\sim$0.3 kg,
with an energy threshold below 0.4~keVee and fully built with previously selected radiopure materials.
This work focuses on the commissioning of the actual setup situated in a laboratory on surface.
A preliminary background model of the experiment is also presented, based on Geant4 simulations
and two discrimination methods: a conservative muon/electron and one based on a $^{252}$Cf source.
Based on this model, TREX-DM could be competitive in the search for low mass WIMPs and,
in particular, it could be sensitive to the WIMP interpretation of the DAMA/LIBRA hint.
\end{abstract}

\section{Motivation}
The main strategy of Dark Matter experiments \cite{Baudis:2012lb} is based on accumulating large target masses
of heavy nuclei (like Xenon), keeping low background levels by a systematic radiopurity control of all components
and an enhancement of the electron/neutron discrimination methods.
However, some recent positive hints, which may be interpreted in terms of low mass WIMPs, 
have changed the detection strategy to sub-keV energies and light gases.
This research line could be led in future experiments by Time Projection Chambers (TPCs),
as they can reach energy thresholds $\sim$ 100 eV and have access to richer topological information.
In contrast to current gaseous-based experiments, focused on directional Dark Matter detection \cite{Ahlen:2013sa},
the TREX-DM experiment proposes a strategy based on high gas pressures,
even if neutron/electron discrimination could be less effective, but keeping a low energy threshold.
TREX-DM is a low background Micromegas-based TPC for low-mass WIMP detection and will profit from all developments
made in Micromegas technology \cite{Giomataris:2006yg, Andriamonje:2010sa},
as well as in the selection of radiopure materials \cite{Cebrian:2011sc, Aznar::2013fa},
specially in CAST \cite{Aune:2014sa} and NEXT-MM \cite{Alvarez:2014va} projects.
Its main goal is the operation of an active detection mass $\sim$0.3 kg
with an energy threshold below 0.4~keVee (as already observed in \cite{Aune:2014sa}).

\section{Description and comissioning}
\begin{figure}[htb!]
\centering
\includegraphics[width=88mm]{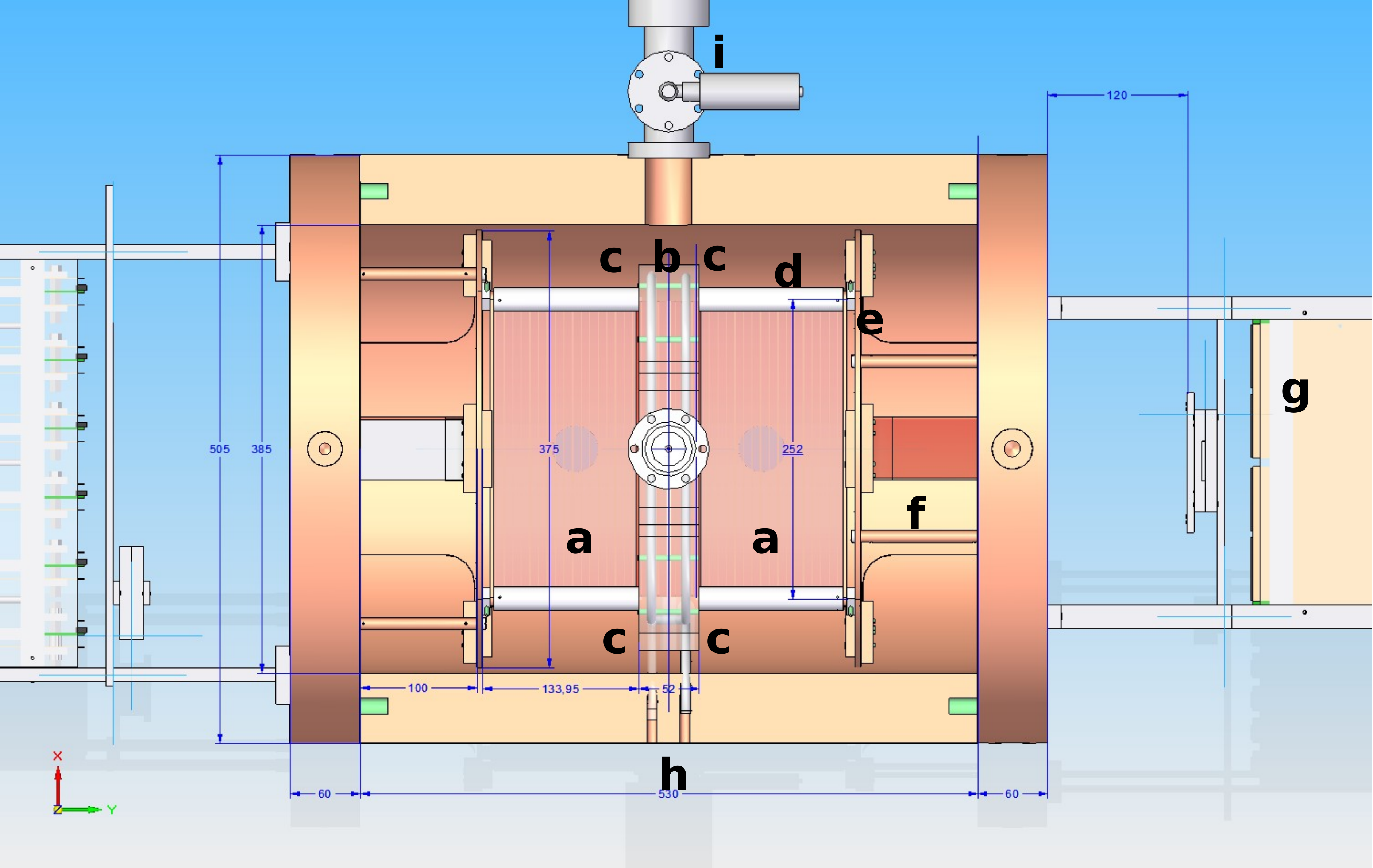}
\includegraphics[width=54mm]{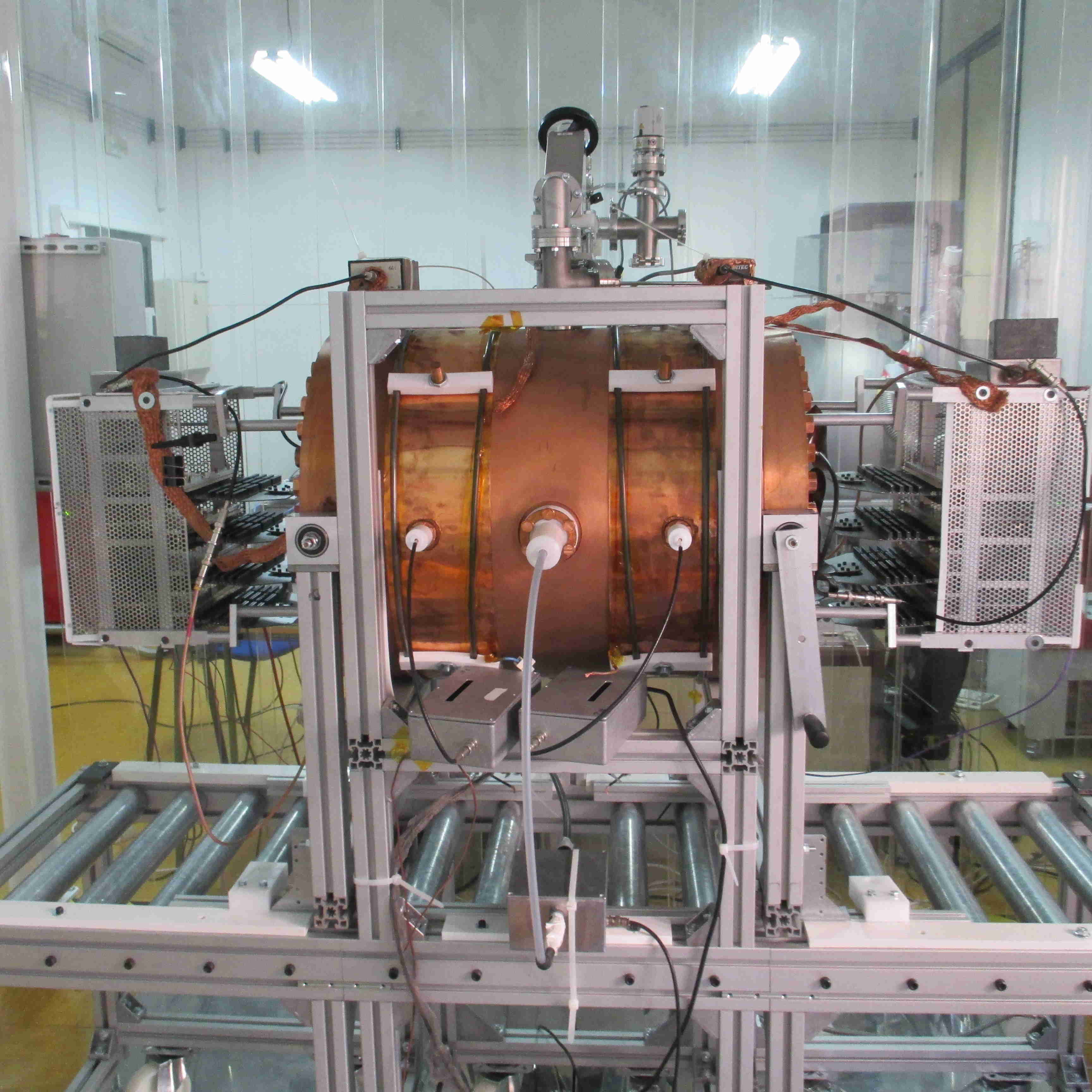}
\caption{Left: Design of the TREX-DM detector. Its different parts are described in detail in the text:
active volumes (a), central cathode (b), calibration points (c), field cage (d), Micromegas detector and support base (e),
flat cables (f), AFTER-based electronics (g), gas system (h) and pumping system (i). Right: A view of the experiment
during the comissioning.}
\label{fig:Setup}
\end{figure}

The actual setup (Fig. \ref{fig:Setup}) is composed of a copper vessel, with an inner diameter of 0.5~m,
a length of 0.5~m and a wall thickness of 6~cm.
The vessel contains two active volumes ({\it a} in the design), separated by a central copper cathode ({\it b}). At each side
there is a field cage ({\it d}) that makes uniform the drift field along the 19~cm between the cathode and the detector.
Each bulk Micromegas detector ({\it e}) \cite{Iguaz:2011fa} is screwed to a copper base,
which is then attached to the vessel's inner walls by means of
four columns. The gas enters the vessel by a feedthrough at the bottom part ({\it h}) and comes out by another one
at the top part ({\it i}). The calibration system consists of a plastic tube entering in the bottom part ({\it h}), which
allows to calibrate each side at four different points ({\it c}) with a $^{109}$Cd source,
emitting X-rays of 22.1 (K$_\alpha$) and 24.9 keV (K$_\beta$).

\medskip
The TREX-DM prototype is part of the wider scope ERC-funded project called TREX, that since 2009 is devoted to
R\&D on low background TPCs and their potential applications in axion, double beta decay and dark matter experiments.
Work on the TREX-DM prototype started 2012 with the first designs and it is now being commissioned at the TREX lab at Zaragoza.
Most of the components have been validated:
the leak-tightness of all feedthroughs has been verified for pressures up to 10 bar;
the drift cage has been tested at high voltage;
and all experimental parameters like the pressure, the temperature and voltages are continuously monitored by a slow control.
Moreover, during the first semester of 2015, some issues have been successfully solved:
the noise level has been effectively reduced by a new High Voltage filter for the central cathode
and a Faraday cage for the interface cards;
a new field cage has been installed to reduce border effects;
and a new DAQ to read both detectors at a rate of 45~Hz each side has been installed.
During the next months, the detector will be characterized in Ar+2\%iC$_4$H$_{10}$ and Ar+5\%iC$_4$H$_{10}$,
with the aim to detect sub-keV energies at high gas pressures.
In parallel, the first designs of a fully radiopure setup are being made,
which include a lead shielding and the replacement of some dirty components in terms of radiopurity.

\section{Background model of TREX-DM}
The sensitivity of the experiment has been studied creating a first background model,
if it were installed at Canfranc Underground Laboratory (LSC).
We have considered two light gas mixtures at 10 bar: Ar+2\%iC$_4$H$_{10}$ and Ne+2\%iC$_4$H$_{10}$;
with an active mass of 0.3 and 0.16~kg respectively and which are good candidates to detect low mass WIMPs.
However, the sensitivity of an argon-based mixture
may be limited by one of its isotopes ($^{39}$Ar), which is $\beta$-decay and has a long life-time.
In our model, we have considered the lowest content of this isotope, measured
in argon extracted from undeground sources \cite{Xu:2015jx}.
We have also simulated the main radioactive isotopes
of all the inner components using their measured activities \cite{Cebrian:2011sc, Aznar::2013fa}
and the cosmic muon flux in Canfranc. In some cases like the Micromegas detectors
or their connectors, we have considered the activities of their radiopure alternative.
The external gamma flux has not been included as its contribution may be supressed by an external shielding.

\begin{figure}[htb!]
\centering
\includegraphics[width=80mm]{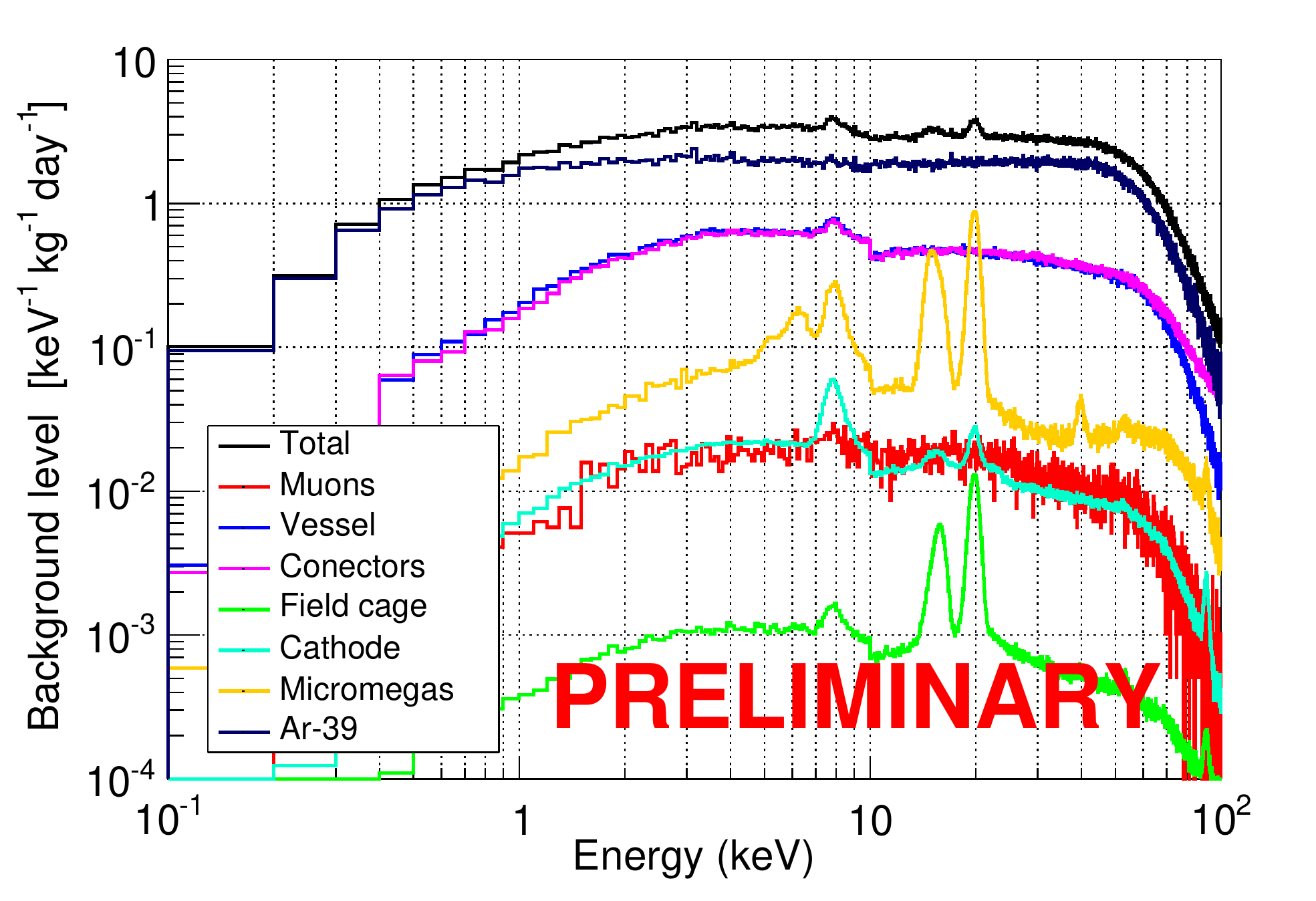}
\includegraphics[width=62mm]{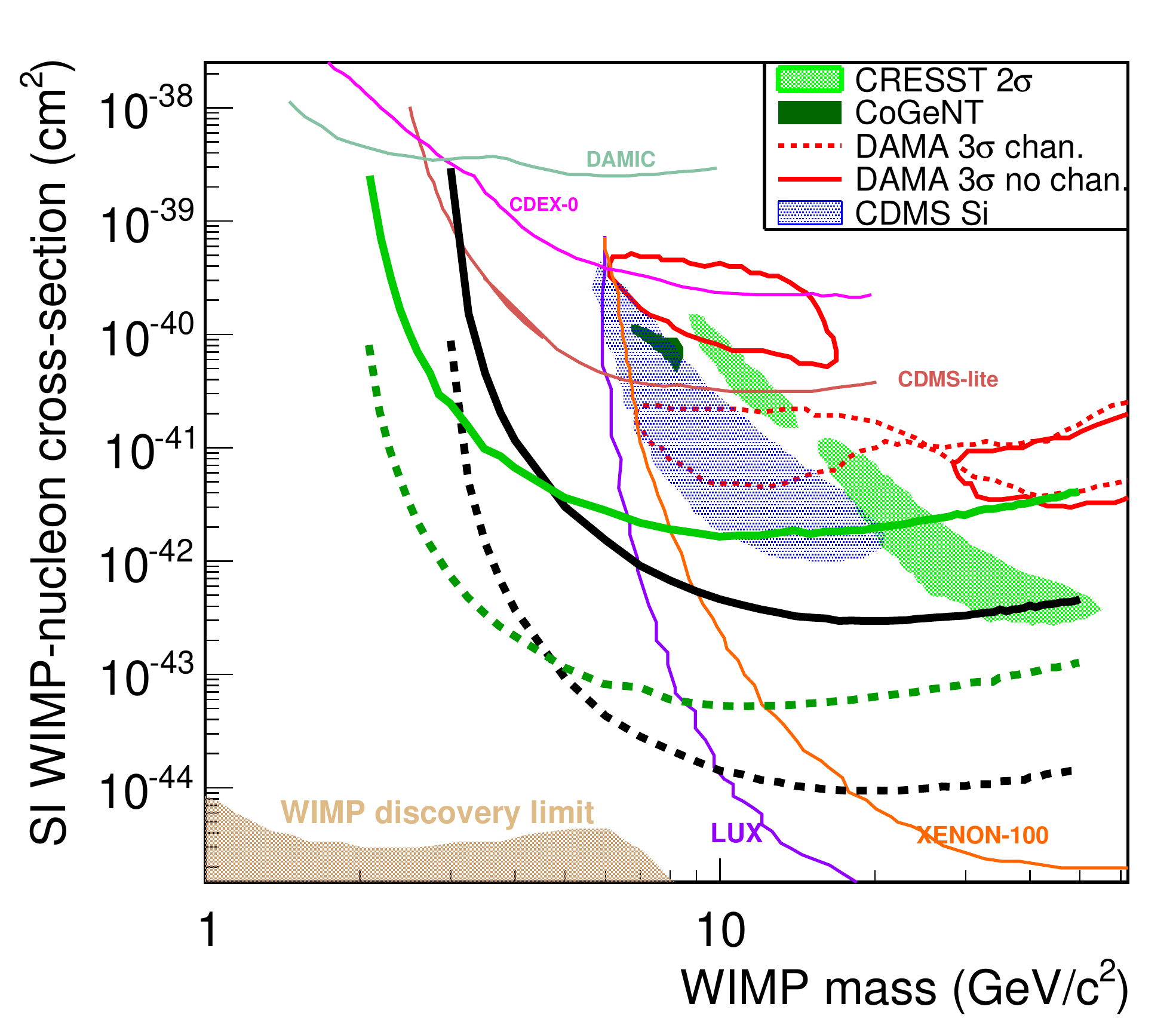}
\caption{Left: Background spectrum expected in TREX-DM experiment (black line) during a physics run
in an underground laboratoy if operated in Ar+2\%iC$_4$H$_{10}$ at 10 bar.
The contribution of the different simulated components is also plotted: external muon flux (red line),
vessel contamination (blue line), connectors (magenta line), field cage (green line), central cathode (brown line),
Micromegas detector (purple line) and $^{39}$Ar (dark blue line).
Right: WIMP parameter space focused on the low-mass range. Filled regions represent the values that may explain
the hints of positive signals observed in CoGeNT, CDMS-Si, CRESST and DAMA/LIBRA experiments. The thick lines are
the preliminary sensitivity of TREX-DM supossing a 0.4~keVee energy threshold and two different hypothesis
on background and exposure: 100 (solid) and 1(dashed) keV$^{-1}$ kg$^{-1}$ day$^{-1}$,
and 1 and 10 kg-year respectively, and for both argon- (black) and neon-based mixtures (green).}
\label{fig:BackSpec}
\end{figure}

\medskip
Two analysis have been used in this background model.
The first one is a modified version of CAST \cite{Aune:2014sa}, optimized to
discriminate low energy X-rays from complex topologies like gammas and cosmic muons.
It uses two likelihood functions generated by the X-rays' cluster features of a calibration source.
Fixing a total 80\% signal efficiency, the expected background level for an argon- (neon-) based mixture gas at 10 bar
is $\sim$3.1 ($\sim$1.4) keV$^{-1}$ kg$^{-1}$ day$^{-1}$, dominated by the $^{39}$Ar isotope in the case of argon and by
the connectors and the vessel in the case of neon.
The contribution of each component is shown in Fig. \ref{fig:BackSpec} (left) for the argon case.
The second analysis is based on the simulation of a $^{252}$Cf neutron source, which reproduces better WIMPs signals.
The level obtained in argon is a $\sim$44\% lower, as nuclear recoils show narrower cluster widths.
Suposing a 0.4~keVee energy threshold and former background levels,
TREX-DM experiment could be sensitive to a relevant fraction of the low-mass WIMP parameter space
(see Fig. \ref{fig:BackSpec}, right) including the regions invoked in some interpretations of
DAMA/LIBRA results and other hints of positive WIMPs signals, with an exposure of 1 kg-year. 

\section{Conclusions and prospects}
The TREX-DM is a low background Micromegas-based TPC for low-mass WIMP detection. Its main goal is the operation of
a light gas at high pressure (active mass $\sim$0.3 kg) with an energy threshold of 0.4~keVee or below
and fully built with previously selected radiopure materials. The detector is being comissioned at TREX laboratory 
and may be installed at the LSC during 2016 for a possible physics run.

\section*{Acknowledgments}
We acknowledge the Micromegas workshop of IRFU/SEDI and the
Servicio General de Apoyo a la Investigaci\'on-SAI of the University of Zaragoza.
We acknowledge the support from the European Commission under the European Research Council T-REX Starting Grant ref.
ERC-2009-StG-240054 of the IDEAS program of the 7th EU Framework Program.
We also acknowledge support from the Spanish Ministry MINECO under contracts ref. FPA2008-03456 and
FPA2011-24058, as well as under the CPAN project ref. CSD2007-00042 from the Consolider-Ingenio 2010 program.
These grants are partially funded by the European Regional Development funded (ERDF/FEDER).
F.I. acknowledges the support from the Juan de la Cierva program and
T.D. from the Ram\'on y Cajal program of MICINN.

\begin{footnotesize}

\end{footnotesize}

\begin{thebibliography}{99}
\bibitem{Baudis:2012lb}
L.~Baudis,
``Direct dark matter detection: The next decade'',
{\it Physics of the Dark Universe} {\bf 1}, 94-108 (2012).
\bibitem{Ahlen:2013sa}
S.~Ahlen {\it et al.},
``The Case for a Directional Dark Matter Detector and the Status of Current Experimental Efforts'',
{\it Int. Jour. Mod. Phys. A} {\bf 25}, 1 (2010).
\bibitem{Giomataris:2006yg}
I.~Giomataris {\it et al.}
''Micromegas in a bulk``,
{\it Nucl. Instrum. Meth. A} {\bf 560}, 405 (2006).
\bibitem{Andriamonje:2010sa}
S.~Andriamonje, D.~Attie, E.~Berthoumieux, M.~Calviani, P.~Colas {\it et al.},
``Development and performance of Microbulk Micromegas detectors'', 
{\it JINST} {\bf 5}, P02001 (2010).
\bibitem{Cebrian:2011sc}
S. Cebri\'an {\it et al.},
``Radiopurity of micromegas readout planes'',
{\it Astropart. Phys.} {\bf 34}, 354 (2011).
\bibitem{Aznar::2013fa}
F.~Aznar {\it et al.}
``Assesment of material radiopurity for Rare Event experiments using Micromegas'',
{\it JINST} {\bf 8}, C11012 (2013).
\bibitem{Aune:2014sa}
S.~Aune, J.~Castel, T.~Dafni, M.~Davenport, G.~Fanourakis {\it et al.},
``Low background x-ray detection with Micromegas for axion search'',
{\it JINST} {\bf 9}, P01001 (2014).
\bibitem{Alvarez:2014va}
V.~Alvarez {\it et al.}
``Description and commissioning of NEXT-MM prototype: first results from operation in a Xenon-Trimethylamine gas mixture'',
{\it JINST} {\bf 9}, P03010 (2014).
\bibitem{Iguaz:2011fa}
F.J.~Iguaz {\it et al.},
``Micromegas detector develpments for Dark Matter directional detection with MIMAC'',
{\it JINST} {\bf 6}, P07002 (2011).
\bibitem{Xu:2015jx}
J.~Xu {\it et al.},
``A study of the trace $^{39}$Ar content in argon from deep undergound sources'',
{\it Astr. Part.} {\bf 66}, 53 (2015).
\end{thebibliography}
\end{document}